\documentclass[12pt]{iopart}
\usepackage{graphicx,color,amssymb,amsfonts,hyperref}
\newcommand{\beq}{\begin{equation}}
\newcommand{\eeq}{\end{equation}}
\newcommand{\bqa}{\begin{eqnarray}}
\newcommand{\eqa}{\end{eqnarray}}
\newcommand{\nn}{\nonumber}

\newcommand{\dbd}[1]{\frac{\partial}{\partial {#1}}}
\newcommand{\rt}[1]{\sqrt{#1}\,}
\newcommand{\erf}[1]{Eq.~(\ref{#1})}
\newcommand{\erfs}[1]{Eqs.~(\ref{#1})}

\newcommand{\bra}[1]{\left\langle{#1}\right|}
\newcommand{\ket}[1]{\left|{#1}\right\rangle}

\newcommand{\sq}[1]{\left[ {#1} \right]}
\newcommand{\cu}[1]{\left\{ {#1} \right\}}
\newcommand{\ro}[1]{\left( {#1} \right)}
\newcommand{\an}[1]{\langle{#1}\rangle}

\newcommand{\st}[1]{\left| {#1} \right|}

\renewcommand{\d}{{\rm d}}

\newcommand{\red}[1]{#1}

\newtheorem{theorem}{Theorem}

\newtheorem{exercise}[theorem]{Exercise}

\begin{document}

\title[]{How many principles does it take to change a light bulb \ldots\ into a laser?}
\author{Howard M. Wiseman}
\address{
Centre for Quantum Dynamics, Griffith University, Brisbane, Queensland 4111, Australia}

\begin{abstract}
Quantum optics did not, and could not, flourish without the laser. The present paper is not about the principles of laser construction, still less a history of how the laser was invented. Rather, it addresses the question: what are the fundamental features that distinguish laser light from thermal light? 
The obvious answer, ``laser light is coherent'', is, I argue, so vague that it must be put aside at the start, albeit to revisit later. A more 
specific, quantum theoretic, version, ``laser light is in a coherent state'', is simply wrong in this context: 
both laser light and thermal light can equally well be described by coherent states, with amplitudes that vary stochastically in space. Instead, my answer to the titular question is that four 
principles are needed: high directionality, monochromaticity, high brightness, and stable intensity. Combining 
the first three of these principles suffices to show, in a quantitative way --- involving, indeed, very large dimensionless quantities (up to $\sim10^{51}$) --- that a laser must be constructed very differently from a light bulb. This quantitative analysis is quite simple, and is easily relatable to ``coherence'',  
yet is not to be found in any text-books on quantum optics to my knowledge. 
The fourth principle is the most subtle and, perhaps surprisingly, is the only one related to 
coherent states in the quantum optics sense:  it implies  
that the description in terms of coherent states is 
the {\em only} simple description of a laser beam. Interestingly, this leads to the (not, as it turns out, entirely new) prediction 
that narrowly filtered laser beams are indistinguishable from similarly filtered thermal beams.   
I hope that other educators find this material useful; it may contain surprises even for researchers who have been in the field longer than I have. \end{abstract}

\maketitle

\section{Introduction} \label{sec:intro}

The importance of the laser to quantum optics, and to optical sciences and technologies more generally, can hardly be over-stated. But what makes laser light special? More specifically, what differentiates it from thermal radiation, which was mankind's only source of illumination for most of history? A one-word answer that springs to the mind of many is `coherence'.  Indeed, wikipedia~\cite{wikiLaser} \red{states}
\begin{quote}A laser differs from other sources of light in that it emits light coherently.
\end{quote} 
For those with a quantum optics education, a more technical answer may spring to mind: laser light is in a coherent state. This idea has long~\cite{PicWil65} been attributed to Glauber, although what he actually said in 1963~\cite{Gla63} (when the term `laser' had yet to supplant `optical maser') is much more nuanced: 
\begin{quote}
The density operator which represents an actual maser beam is not yet known. It is clear that such a beam cannot be represented by \dots\ [a] coherent state \dots\  unless the phase and amplitude stability of the device is perfect. On the other hand, a maser
beam is not at all likely to be described by \dots\ [an]  incoherent classical model \dots. More plausible models for a steady maser beam are much closer in behavior to the 
ideal coherent states.
\end{quote}

In this pedagogical paper I maintain that the word `coherent' has too many meanings to be a useful answer in itself, while the idea that a laser is in a coherent state is simply wrong in the context of trying to understand how laser light differs from thermal light. Instead, I will argue that there are four key features which distinguish laser light from thermal light. This paper is based upon material I teach to advanced undergraduate students as a small part of an introductory course on  quantum electrodynamics. In class I ask them the question ``what makes laser light special?''  
and invariably educe with little difficulty three of the four key features I have in mind: 
\begin{enumerate}
\item High directionality;
\item Monochromaticity;
\item High brightness.
\end{enumerate}
Pleasingly, these three are also easily drawn out  from the following statement on the official {\em Year of Light} home page~\cite{iyolLaser}
\begin{quote}
A laser is an optical amplifier --- a device that strengthens light waves.  Some lasers have a {\em well-directed}, {\em very bright} beam with a {\em very specific color}\,; others emphasize different properties, such as extremely short pulses.  The key feature is that the amplification makes light that is very {\em well defined} and reproducible, unlike ordinary light sources such as the sun or a lamp.
\end{quote}
Ignoring the clause covering lasers that are not CW, my first three key features of laser light are clearly stated here (albeit in a different order), as emphasised in italic font by me.  
I have also emphasised in the above quote ``well defined'' because, as we will see, there is a 
property of laser light that is well-defined, and not covered by the above three features:  
\begin{enumerate}
\item[(iv)] Stable intensity.
\end{enumerate}

The above four criteria were already formulated as ``four quantitative conditions that the output of a device must satisfy in order for the device to be considered a laser'' by me in 1997~\cite{Wis97}.  What is new 
in the present paper is two-fold. First, I give a detailed and quantitative 
examination of how light from a light bulb fails to satisfy these criteria, and how,  
even using spatial and frequency filtering, it would be wildly impractical to obtain a beam 
satisfying the first three criteria, and impossible to create one satisfying all four. 
Second, I discuss why that which Glauber implied was unknown (in 1963) --- an explicit 
expression for the density operator which represents an actual laser beam, written in a 
way comparable to that for the density operator for thermal radiation ---  is still unknown 
today. That is not to say that the state of a laser beam cannot be described; 
\red{the quantum properties of a laser beam have been understood since the 
later 1960s (see the textbooks~\cite{Lou73,SarScuLam74} and references therein).  
But this is not the same as an explicit equation for the multimode density operator, as can be given 
for the case of thermal light.}
Interestingly, this difference has 
experimental consequences for the intensity fluctuations of frequency-filtered laser light, 
as I will discuss. 

The remainder of this paper is organised as follows. I begin in Sec.~\ref{sec:mix} by introducing some notation and identities for single-mode mixed states that will be useful subsequently. Then, in Sec.~\ref{sec:bulb},  
I consider the multimode 
mixed state describing emission from a light bulb. This leads to the discussion, in Sec.~\ref{sec:props}, as to what could or could not be done to light-bulb light to give it the above four   properties of laser light. In Sec.~\ref{sec:coll} I consider collimation to achieve directionality, and calculate how hot the light-bulb filament would need to be to give a beam of the same power as a laser beam. In Sec.~\ref{sec:mono} I consider spectral filtering to achieve (near) monochromaticity, and again calculate the required light-bulb temperature to reproduce a beam with the power and spectral properties of a laser. In Sec.~\ref{sec:bright} I show how it is the beam brightness that results in the astronomically high temperatures calculated in the preceding sections, and also calculate what fraction of the radiated power would survive the required spatial and frequency filtering. 
The staggeringly small answer gives, I think, the best feel for how special laser light is. In Sec.~\ref{sec:stable} I explain how, even if all of this were possible, it would still not give a beam satisfying the fourth criterion. 
While the difference is simple to state in the time (or longitudinal position) domain, the contrast 
is even more marked in the frequency (or wavenumber) domain: there is a simple expression for the 
quantum state of thermal light in terms of frequency components, but no such simple expression 
for the state of laser light. This last topic --- the quantum state of a laser beam --- does require something beyond undergraduate-level mathematics to understand, and so most of the analysis  is presented in the Appendix. I conclude in Sec.~\ref{sec:conc} with a reconsideration of the notion of `coherence', and a return to the Glauber quote from the opening paragraph. 

\section{Single Mode Mixed States} \label{sec:mix}

Two types of single-mode mixed state will play a role in the later sections of this paper. 
The first is a thermal state. For a harmonic 
oscillator, such as a single mode of the EM-field, in thermal equilibrium at 
temperature $T$, the mean number of excitations (photons) is 
\beq \label{defnth}
\bar{n}_{\rm th}(\omega)= \frac{1}{\exp(\hbar\omega/k_B T)-1}.
\eeq
In terms of this, the state matrix, also known as a density operator, 
for a single-mode thermal (SMT) state can be written as 
\beq \label{smt:num}
\rho_{\rm th}\left(\bar{n}\right)
= \frac{1}{1+\bar{n}}\sum_{n=0}^\infty \left(\frac{\bar{n}}{\bar{n}+1}\right)^n
\ket{n}\bra{n}.
\eeq
The photon-number variance of the SMT state is $\bar{n}^2 +\bar{n}$. That is, 
the standard deviation in $n$ is always larger than the mean. 
The SMT state can also be expressed in terms of coherent states as
\beq \label{SMT:coh}
\rho_{\rm th}(\bar n) = \frac{1}{\pi \bar{n}} \int \d^2\alpha\, 
\exp(-|\alpha|^2/\bar{n}) \ket{\alpha}\bra{\alpha}.
\eeq

The second type of single-mode mixed state is what I will call a 
single-mode laser (SML) state. This is meant to represent the state 
of the cavity mode for a laser in steady state.  An ideal laser, far above threshold, 
will have a near-Poissonian photon number distribution, the same 
as a coherent state~\cite{Wis97,Lou73,SarScuLam74,Lou83}. But the phase of the laser, 
relative to any other optical-frequency clock\footnote{See Ref.~\cite{Wis04a} 
for arguments as to why a laser is as much a clock as anything can be at optical frequencies.} 
 will be completely 
unpredictable. Even if the phase $\phi$ were established at some time, 
it would quickly (on a human time-scale; often slowly on an atomic physics 
time-scale) become undefined, through a process of phase diffusion. The simplest model for this, 
yielding a laser with a Lorentzian spectrum, is a constant-diffusion Fokker--Planck equation 
for the phase probability distribution~\cite{Wis97,Lou73}: 
\beq \label{phasebrown}
\dbd{t}\,p(\phi) = \frac{\Gamma}{2} \ro{\dbd{\varphi}}^2 p(\phi),
\eeq
where $\Gamma$ is the laser linewidth, the reciprocal of its coherence time.  

Thus if a SMT state is described by the mixture (\ref{SMT:coh}), 
a SML state is described by the mixture 
\beq \label{mixeq1}
\rho_{\rm laser}(\mu) = \frac{1}{2\pi}\int \d\phi 
\ket{\sqrt{\mu}\,e^{i\phi}}\bra{\sqrt{\mu}\,e^{i\phi}}.
\eeq
Here $\ket{\sqrt{\mu}\,e^{i\phi}}$ is a coherent state, with a mean photon number 
of $\mu$. Equivalently, the SML state can be written 
\beq \label{mixeq}
\rho_{\rm laser}(\mu) = \sum_{n=0}^\infty e^{-\mu}\frac{\mu^n}{n!} 
\ket{n}\bra{n}.
\eeq
There is no more reason to say that a SML is `really' in a coherent state 
than that a SMT state is. The crucial difference between them is that in a SML 
the standard deviation in the photon number 
is much less than the mean $\mu$, as long as the latter is large (as it will typically be).

\section{Light-bulb light} \label{sec:bulb} 

Thermal light is the sort of light that comes out an old-fashioned 
(incandescent) light bulb. If we use $\kappa$ as an index for the modes --- 
radiating in all directions, with all frequencies, and with both polarizations --- 
populated by light from the light bulb, the total state of these modes 
will simply be a tensor product over all the modes: 
\beq
\rho = \bigotimes_\kappa \rho_{\rm th}^\kappa\left(\bar{n}_{\rm th}(\omega_\kappa)\right). 
\eeq
Here $\bar{n}_{\rm th}(\omega_\kappa)$ is as defined in \erf{defnth}, and 
$\rho_{\rm th}$ is the SMT state of \erf{smt:num}. 
This tensor-product form follows simply from the fact that the energy is additive over all these modes (each one is a harmonic oscillator). 

The Stefan--Boltzmann law says that 
a blackbody source of area $A$, at temperature $T$, has a radiative power of 
\beq \label{SBL}
P_{\rm total} = \frac{\pi^2}{60} \frac{A\,(k_B T)^4}{c^2\hbar^3}.
\eeq
Wien's displacement law says that 
$\lambda_{\rm max}$, the wavelength of maximum spectral power, is given by 
\beq \label{WDL}
\lambda_{\rm max} =  \frac{2\pi \hbar c}{x k_B T}\;;\;\; x \approx 4.965.
\eeq
Combining these, one finds 
\beq
A \approx 2.37\, c^{-2}\, \hbar^{-1} P \,\lambda_{\rm max}^4
\eeq
We can use this to calculate the effective area of the radiating element of a light-bulb, 
using the fact that the spectrum of a typical light bulb has its peak power in the infra-red, with $\lambda_{\rm max} \approx 1$ micron. For a 60 Watt bulb this gives $A\approx 15$ (mm)$^2$, 
which seems reasonable.  This will be used in Sec.~\ref{sec:bright}. 

The rest of this paper is devoted to teasing out 
how the thermal light described here is different from laser light.


\section{Changing Light Bulb Light into Laser Light} \label{sec:props}

\subsection{Collimated Polarized Thermal Light} \label{sec:coll} 

The first obvious property of laser light which 
contrasts with thermal light is that it is not isotropic. Rather, it 
propagates in a single direction, and is usually polarized. 
These properties could, in principle, be produced from a thermal source  
(a light-bulb) by collimating it --- that is, passing it through a series of 
finite apertures and lenses --- and then 
passing it through a polarizer. This will lead, in the 
ideal limit, to a field with modes described by a single parameter, 
the `wavenumber' $k>0$  in the direction of propagation. Since the collimation does 
nothing but exclude some modes, the state of the collimated polarized field is 
still a tensor product of SMT states: 
\beq
\rho = \bigotimes_k \rho^k_{\rm th}\left(\bar{n}_{\rm th}(ck)\right). 
\eeq


Because the light is collimated, the energy density (which scales as 
$(k_B T)^4/(\hbar c)^3$ for the field inside a black-body oven) is not a 
sensible quantity to consider. The energy per unit volume can be changed simply 
by expanding or focussing the collimated beam. Rather, we should calculate the energy 
per unit {\em length} in the direction of propagation. Introducing a 
normalization length $L$, this is given by 
\beq
\frac{\an{\hat H}}{L} = \frac{1}{L} \sum_k \hbar\, ck \, \bar{n}_{\rm 
th}(ck) = \frac{\hbar c}{2\pi} \sum_k (\Delta k) \,k\, \bar{n}_{\rm th}(ck),
\eeq
where $\Delta k = 2 \pi/L$ is the separation of the modes in $k$-space. In the 
limit $L \to \infty$ this separation becomes infinitesimal and the 
sum can be converted to an integral, $\sum_k (\Delta k) \to \int_0^\infty \d k$. 
The result is ${\pi (k_{B}T)^2}/\ro{12\hbar c}$. 

Since the light is propagating in one direction, it is natural to 
convert this result into the power (energy per unit time) in the 
collimated beam:
\beq \label{powcoll}
P_{\rm coll} = \frac{\an{\hat H}c}{L} = \frac{\pi}{12} \frac{(k_{B}T')^2}{\hbar}.
\eeq
This can now be easily compared to the power of the output of a 
laser, which is typically of order 100 milliW. Solving for the temperature, 
one finds $T' = 4.6 \times10^5$ K. Here we see the reason for using $T'$ rather 
than $T$: the required temperature is very different from that of an actual light bulb, 
 $T \approx 3000$ K. Indeed, to replicate the laser power, our  
``light bulb'' would have to be much hotter than the surface of the sun, and indeed hotter than the surface of the hottest known stars (newly formed white dwarfs)~\cite{WhiteDwarf}.

I ask students to perform the above calculation, and then ask them in class what conclusions they draw from this. The answer is usually readily forthcoming: that a half-million-degree light bulb is not what is actually hiding inside a typical laser, so a laser must be different from a thermal source.

\subsection{Monochromatic Collimated Polarized Thermal Light} \label{sec:mono}

As well as being unidirectional and polarized, laser light is close 
to monochromatic. That is, almost all of the power is in a narrow frequency band. 
This property could be achieved from 
collimated polarized thermal light by passing it through filters of increasingly 
narrow frequency resolution (such as a refracting prism, then a diffraction grating, then a series of Fabry-Perot etalons). The overall transmission can be described by a filter function $0 \leq f(\omega)\leq1$.  To mimic the frequency spread of a typical laser, 
we can take the filter function to be  Lorentzian, with 
\beq \label{deffom}
f(\omega) = \frac{(\Gamma/2)^2}{(\Gamma/2)^2 + 
(\omega - \omega_0)^2},
\eeq
where $\omega_0$ is the mean frequency and $\Gamma \ll \omega_{0}$ is the 
FWHMH (Full Width at Half Maximum Height) linewidth. 
Note that for a laser, this $\Gamma$ is 
the phase diffusion rate mentioned in Sec.~\ref{sec:mix}. 
Since this filtering is a passive process, the state of each mode $k$ remains 
a SMT state, just with a modified mean occupation number, 
\beq
\bar{n}_{\rm f}(\omega) = \bar{n}_{\rm th}(\omega)f(\omega) 
\simeq {\nu}\, f(\omega) , \label{filteredn}
\eeq
where ${\nu} \equiv \bar{n}_{\rm th}(\omega_0)$. 
Here the approximation holds in the frequency band of interest --- around $\omega_0$ --- 
although not for very low (radio) frequencies $\lesssim \Gamma$.   In the optical band, the state of the 
filtered light can thus be written as 
\beq \label{cpfts}
\rho = \bigotimes_k \rho^k_{\rm th}\left({\nu}\, f(\omega)\right).
\eeq

The power in this filtered collimated polarized thermal light is 
\beq
\frac{\an{\hat H}c}{L} = \frac{c}{L} \sum_k \hbar\, ck\, \bar{n}_{\rm f}(ck).
\eeq
In the limit $\Gamma \ll \omega_{0}$ this can be evaluated as
\beq \label{powfil1}
P_{\rm filt} = {\nu}\, \hbar \omega_0 \, \Gamma / 4
\eeq
For high temperatures ($k_B T \gg \hbar \omega_0$, which, unsurprisingly, will be the 
relevant limit) the expression for ${\nu} = \bar{n}_{\rm th}(\omega_0)$ simplifies to give 
\beq \label{powfil2}
P_{\rm filt} = k_B T''\, \Gamma / 4
\eeq
I use $T''$ to emphasize that this is a different temperature both from an actual 
light-bulb temperature $T$ and from that required in the preceding section, when considering 
only collimation, $T'$. Let us compare with a laser of moderate quality, with an output 
power of order 100 milliW (as above) and a linewidth $\Gamma$ of order $10^7$ 
s$^{-1}$ (1.6 MHz). This time, the required temperature is $T'' = 2.9\times10^{15}$ K. 
This is far higher than any temperature that has been produced on the earth, and corresponds 
to that of the universe when it was less than $10^{-12}$ seconds old. 
This time the message is even more emphatic: a laser is profoundly different 
from a thermal source.

\subsection{Bright Monochromatic Collimated Polarized Thermal Light} \label{sec:bright}

The reason such astronomically high temperatures are required for a hypothetical 
thermal source behind a laser beam is that so much of the original radiation would have to be discarded 
to obtain a beam with the desired properties. One can learn more by examining just 
how much light must be thrown away. 

First, consider the collimation process. Using \erfs{SBL}, (\ref{WDL}), and (\ref{powcoll}), 
one can show that, by numerical coincidence, the proportion of 
power {\em not} discarded in the necessary collimation process is 
extremely well approximated by a simple ratio: 
\beq \label{beauty}
\frac{P_{\rm coll}}{P_{\rm total}} \approx \frac{\lambda'_{\rm max}{}^2}{A}
\eeq
Here $\lambda_{\rm max}'$ is the peak wavelength corresponding to temperature $T'$ 
defined in \erf{powcoll}.  
This scaling is easily understood from elementary transverse coherence theory. If one considers a large sphere of radius $R$, concentric with a source of size $r$ that produces light of wavelength $\lambda$, the coherence length on the surface of the large sphere scales as $R\lambda/r$~\cite{Hecht01}. Hence the number of transversely coherent modes in the far field scales as $(\lambda/r)^2$, so discarding all but one gives the type of ratio appearing in \erf{beauty}. For the parameters used in Sec.~\ref{sec:coll}, and using $A=15$ (mm)$^2$ from Sec.~\ref{sec:bulb}, \erf{beauty} evaluates to $2.6\times10^{-12}$. 

Next, consider adding filtering. This time, we will ignore factors of order unity, for simplicity. 
Then, using \erfs{SBL}, (\ref{WDL}), and (\ref{powfil1}), we find 
\beq \label{truthiness}
\frac{P_{\rm filt}}{P_{\rm total}} \sim \frac{\lambda''_{\rm max}{}^2}{A}\frac{\Gamma}{\omega_{\rm max}''}. 
\eeq
Here $\lambda_{\rm max}''$ and $\omega_{\rm max}''$ refer to the temperature $T''$ 
defined in \erf{powfil1}. For the 
parameters in Sec.~\ref{sec:mono},  the first (collimation) factor evaluates to something far smaller even than that in the preceding paragraph, this time of order $10^{-31}$, while the second (frequency filtering) factor evaluates to $\sim 10^{-20}$. 
 Thus \erf{truthiness} evaluates to $\sim 10^{-51}$, a dimensionless number 
whose reciprocal could truly be described as astronomical. 

Equation (\ref{truthiness}) involves the parameter $\omega_{\rm max}''$ (or $\lambda_{\rm max}''$) which has no simple interpretation in terms of the properties of the laser beam, unlike $\omega_0=2\pi c  / \lambda_0$, the actual central frequency of the laser. However, using the (appropriate) high-temperature limit $k_{\rm B}T={\nu}\,\hbar\omega_0$, one can rewrite \erf{truthiness} as  
 \beq \label{truth}
\frac{P_{\rm filt}}{P_{\rm total}} \sim \frac{\lambda_{0}^2}{A}\frac{\Gamma}{\omega_0}\frac{1}{{\nu}^3}. 
\eeq 
Despite initial appearances, this, like \erf{truthiness}, is independent of $\omega_0$.   
 To analyse the contribution of each term we must choose a value of $\lambda_0$. Let us take $\lambda_0 = 1$ micron, the same as $\lambda_{\rm max}$ for the original (realistic) light bulb of Sec.~\ref{sec:bulb}, and a 
pretty representative figure for lasers. The first fraction in \erf{truth} is the geometric factor one might naively expect from collimation, as per the argument in the first paragraph of this section. For the above parameters, it evaluates to $\sim 10^{-7}$. The second fraction is a factor one might naively expect from filtering. For the above parameters, it evaluates to $\sim 10^{-8}$.  Thus, the greatest contribution to the overall `efficiency' of $\sim 10^{-51}$ is from the third fraction. For the above parameters, ${\nu} \sim 10^{12}$, so that ${\nu}^{-3} \sim 10^{-36}$, as required. 
 The reason this factor appears is that, in order to achieve the actual brightness of a laser beam, it is necessary to start with an extremely hot source, so that $\lambda_{\rm max}'' \ll \lambda_0$ and $\omega_{\rm max}'' \gg \omega_0$.  
 
One might object that \erf{truth} contains the parameter ${\nu}$ which has not been given any simple interpretation. But in fact it does have one. From this expression
\beq \label{defMthermal}
{\nu} = \frac{P}{\hbar\omega_0}\frac{4}{\Gamma}
\eeq
one can see that ${\nu}$ is 
the number of photons per unit time, multiplied by the coherence time of the beam (ignoring constants of order unity).  
That is, ${\nu}$ is roughly the number of photons that come out coherent with one another, 
before the phase of the beam randomly shifts to some other value. 
Of course strictly it makes no sense to talk of the phase of a beam made of 
single photons, or containing a fixed number of photons, but hopefully the 
preceding sentence conveys meaning intuitively.  Thus we see that ${\nu}$ is  
a natural dimensionless way to quantify the intensity or brightness of the beam, just as 
$\Gamma/\omega_0$ is the natural way to quantify its monochromaticity. 
And it is the fact that ${\nu} \gg 1$ for typical laser parameters that makes  
it astronomically impractical to create laser-like light by collimation and filtering. 
 
\subsection{Laser Light (a Stable Intensity too)} \label{sec:stable}

We have seen that it is not feasible to produce a beam with the same power 
$P$ and linewidth $\Gamma$ and central frequency $\omega_0$ as a typical laser beam 
by collimating, polarizing, and filtering thermal light, as derived from a light bulb. 
Even if it could be done, the result would still not be equivalent to a laser 
beam. The remaining difference is essentially the same as the difference 
between a single mode thermal state and a single mode laser state. 
The former has a very poorly defined photon number, while the latter 
can have a very well-defined photon number. 

In Sec.~\ref{sec:bright}, I talked about the coherence time in terms 
of the random variation of the phase. In fact, a 
filtered collimated polarized thermal beam would have huge intensity fluctuations 
as well as phase fluctuations. They would both occur on the time scale 
of the coherence time, $\Gamma^{-1}$, and in fact both would contribute equally 
to the decay of coherence.  For those familiar with stochastic calculus, 
 \red{and quantum optics, it is possible to be more specific. 
The state (\ref{cpfts}) is equivalent to a spatially stochastic coherent state. That is, 
it is an ensemble where each member is an eigenstate of all 
the annihilation operators $\cu{a_k}$, with eigenvalues $\cu{\alpha_k}$ (coherent-state amplitudes) 
that differ in different member of the ensemble. Specifically, a randomly drawn member of the 
coherent-state ensemble representing (\ref{cpfts})  
can be generated by choosing each $\alpha_k$ as an independent random variable from the probability distribution $[\pi\bar{n}_f(\omega_k)]^{-1}\exp[-|\alpha_k|^2/\bar{n}_f(\omega_k)]d^2\alpha_k$. An alternative way of expressing this is by converting (via the Fourier transform) 
from $k$-space to $x$-space (longitudinal position), yielding, for any given ensemble member, 
a coherent-state amplitude $\alpha(x)$.  
In contrast to $k$-space, the different $\alpha(x)$ coherent-state amplitudes are not independent random 
variables. Rather, as stated above, they are correlated in both amplitude and phase on a time scale 
of $\Gamma^{-1}$. Specifically, a randomly drawn $\alpha(x)$, for all $x$, 
can be generated with the correct statistics by this equation: 
\beq \label{cpftscs}
\alpha(x) = e^{-ik_0 x}\sqrt{\nu}\int_{-\infty}^{x/c} ds\, ({\Gamma}/{2})e^{(\Gamma/2) (t-x/c)} \zeta(t). %
\eeq
Here $\zeta(t)$ is a complex white noise process~\cite{IkeWat89}, 
satisfying ${\rm E}[\zeta(t)^*\zeta(t')]=\delta(t-t')$, 
with all other first and second-order moments vanishing. The coherent-state amplitude $\alpha(x)$ is 
normalized so that ${\rm E}[|\alpha(x)|^2]$} is the photon flux (mean number of photons per unit time).  
This evaluates, as it should, to  $\nu\Gamma/4$. \red{The direction of propagation has been taken to be in the negative $x$ direction; a particular member of the ensemble is not a stationary state, but rather 
 changes in time by propagation at the speed of light so that $\alpha_t(x) = \alpha_{t+\tau}(x+c\tau)$. 
 The quantum state $\rho$ of the whole ensemble is of course stationary, because the {\em statistics} 
 of \erf{cpftscs} are stationary (invariant under displacements of $x$).} 

\red{A laser beam can also be described as an ensemble of coherent states, which can again 
be thought of as a state with a stochastically varying coherent-state amplitude (varying in space at any 
particular time, or in time as it passes by any particular point in space). 
By contrast with a thermal-derived beam, however,} a laser beam has an essentially fixed intensity. 
Ideally it has only Poissonian fluctuations in the number of photons in any given time interval. 
Specifically, in the limit where the state of the  laser cavity mode is given by 
\erf{mixeq}, and its phase fluctuations are described by \erf{phasebrown}, 
the laser beam has a \red{coherent-state amplitude that 
 varies stochastically with position $x$ as 
\beq \label{laserscs}
\alpha(x) = e^{-ik_0 x}\sqrt\frac{\nu\Gamma}{4}\exp\sq{i\int_{-\infty}^{x/c} dt\, \sqrt{\Gamma}\,\xi(t)} .
\eeq
Here} $\xi(t)$ is a real white noise process, satisfying ${\rm E}[\xi(t)\xi(t')]=\delta(t-t')$. 
That is, the amplitude (by which I mean the modulus of the coherent-state amplitude) is constant, but the 
phase is stochastic. Although this is very different from the stochasticity in \erf{cpftscs}, they have the 
same power spectrum 
\beq \label{firstspectrum}
S(\omega) = \int_{-\infty}^\infty d\tau\, e^{i\omega \tau} \red{{\rm E}[\alpha^*(x+c\tau)\alpha(x)]} = \nu\,f(\omega),
\eeq
which, as defined here, is the photon flux per unit frequency, a dimensionless quantity. 
Recall that the filter function $f(\omega)$ as defined in \erf{deffom} is dimensionless, as is 
${\nu}\gg 1$, the number of photons per coherence time. 

The fact that the stochastic equation (\ref{cpftscs}) is equivalent to a very 
simple explicit expression, Eq.~(\ref{cpfts}), for the quantum state $\rho$ of 
thermal-derived beam 
in terms of single-mode thermal states 
tempts one to assume the analogous relation for a laser beam. That is, for 
physicists sufficiently well read to know that a laser cavity mode is not in a coherent state 
of fixed phase, but rather in a state like that of Eq.~(\ref{mixeq1}), nothing is more 
obvious\footnote{A. Steinberg {\em et al.}, semi-personal communication (Facebook, 12th September, 2015)} 
than the following 
\beq \label{QSLB}
{\rm guess:} \hspace{2em} \rho_{\rm laser} = \bigotimes_k \rho^k_{\rm laser}\left({\nu}f(ck)\right)
\eeq
for the quantum state of a laser in $k$-space. If this were true then further 
filtering {\em within} the laser spectrum would produce a state of the same form, 
just with a narrower spectral function $f(\omega)$, as is the case for the thermal-derived 
beam (\ref{cpfts}). 

I succumbed to the above temptation many years ago, and only the finalisation 
of the current paper \red{in the last week before the 
submission deadline (16th September 2015),} has taught me that \erf{QSLB} is {\em false}.  
Not only is \erf{QSLB} not equivalent to \erf{laserscs}; it does not describe a CW beam 
at all. That is, it does not describe a beam of indefinite length, 
such that the statistical properties of the beam are independent of its length and 
invariant under translations along it. Moreover, a laser beam does {\em not} have 
the property that filtering within the laser spectrum would produce a state of the same form. 
Rather, this extra filtering 
would produce a state with a narrower spectrum but with extra intensity noise\footnote{Of course I am assuming a filter cavity independent of the laser cavity. If the filter cavity is locked to the phase 
of the laser cavity (which could be measured relative to some other, more stable oscillator) then the procedure 
performed is not really filtering --- it is creating a more complicated type of laser. See also the 
{\bf Note Added} immediately before the References.}.
In particular, in the limit of a very narrow filter, the state produced would have 
exactly the same properties as the thermal-derived beam (\ref{cpfts}). That is, it would 
have huge intensity fluctuations, with standard deviation the same size as the mean. 
The proofs of all these non-obvious statements are given in the (late written) Appendix. 

One might wonder whether the above difference between thermal-derived light 
and laser light is a big enough deal to list a stable intensity 
as the final key feature of laser light. 
After all, if a laser has large ($\sim \pi$) phase fluctuations on a time 
scale $\Gamma^{-1}$, why would having large (of order the mean) 
intensity fluctuations on the same time scale make the  light qualitatively 
worse? For the purposes of cutting through a thin sheet of metal with 
an ``industrial strength'' laser, intensity fluctuations would presumably 
make no difference.  
The characteristic time of the cutting process is (I presume) much greater than 
$\Gamma^{-1}$, so the intensity fluctuations would average out. 
But in this example even the narrow linewidth of a laser is perhaps not relevant; 
only the high photon flux and single transverse mode necessary for tight focussing would be. 

However the situation is certainly different for scientific applications such as atomic 
physics experiments. There the system (atoms) typically have a coherence 
time much shorter than $\Gamma^{-1}$. The atomic coherence time 
upper bounds the effective time of a single `shot' of the experiment (e.g. a single 
photon emission from an atom). Thus the laser phase (and intensity) is 
effectively constant over a single shot. Moreover, because atomic processes, 
in continuous-wave experiments, are insensitive to the absolute phase of light, 
effectively the same experiment is performed in each independent shot. But to 
gather good statistics, such an experiment is typically performed over a total 
time of order $\Gamma^{-1}$ or longer. Thus if the intensity of the beam 
were to be fluctuating over that time-scale, this would add a great deal of noise 
to the process being investigated, because atomic processes are certainly not 
independent of the intensity of light. The stable intensity of a laser 
enables the experimenter to perform essentially the same deterministic 
operation on the atoms in every shot. 

Equation (\ref{laserscs}) implies that the photon statistics of the beam are Poissonian. 
This means that in any time interval, the uncertainty in the number of photons 
in the beam is equal to the mean number: $\delta N = \sqrt{\bar N}$. Of course 
this is not exactly true for real systems, and for very long time scales (very low frequencies) 
there will always be fluctuations due to technical noise. 
But from arbitrarily short times up to times much 
greater than the coherence time  $\Gamma^{-1}$, a good laser will have 
$\delta N = O(\sqrt{\bar N})$, which is much smaller than the mean if 
the latter is large. Even over longer times, where technical noise dominates, 
it is still the case that $\delta N \ll {\bar N}$. This is to be contrasted with a collimated and 
filtered thermal source. In that case, if one looks at a time interval of order   
$\Gamma^{-1}$ or shorter, 
one would see the super-Poissonian statistics of a SMT state (\ref{smt:num}),  in which $\delta N > \bar{N}$ --- fluctuations that are 
not small relative to the mean.

\section{Conclusion} \label{sec:conc}

We have seen that there are four key differences between laser light (in the CW regime for simplicity) and light-bulb light. To be quantitative, we can consider typical devices as above: a laser with power of 100 mW, a linewidth $\Gamma$ of $10^7$ s$^{-1}$, and a wavelength $\lambda_0$ of 1 micron; and a light bulb with filament area $A=15$ mm$^2$, and a peak-spectral wavelength $\lambda_{\rm max}$ of 1 micron. Thus: 
\begin{enumerate}
\item Laser light is polarized and has a single transverse mode; light-bulb light is unpolarized and is emitted into something like $A/\lambda_{\rm max}^2 \sim 10^7$ transverse modes. 
\item Laser light is monochromatic, with $\Gamma/\omega_0 \sim 10^{-8}$; light-bulb light is broad-spectrum with $\delta\omega \sim \omega_{\rm max}$. 
\item Laser light is intense, with ${\nu} \sim  10^{12}$ photons per coherence time; light-bulb light has $\bar{n}_{\rm th}(\omega_{\rm max}) = 1/x \approx 0.2$ photons per spatio-temporal mode at spectral peak.   
\item Laser light has a stable intensity, with photocount uncertainty 
$\delta N \ll \bar{N}$ over time intervals long enough that $\bar{N} \gg 1$; 
light-bulb light, if collimated and filtered, would have a photocount uncertainty $\delta N$ larger than $\bar{N}$ for time intervals $\lesssim \Gamma^{-1}$. 
\end{enumerate} 

None of the above principles state the difference in terms of laser light being coherent, 
but all of them can be regarded as aspects of coherence. The first two are purely classical 
aspects of coherence: (i) complete transverse coherence; (ii) very high 
longitudinal coherence. The third is 
quantum, in that it involves photon number per coherence time, a quantity 
that is not defined classically. Stated loosely as the requirement of having 
 ``many photons coherent with one another'', it is clearly an aspect of coherence.  
The fact that ${\nu}\gg 1$ is {\em the} reason 
 it is possible to interfere light from two independent 
 lasers, as in Ref.~\cite{PflMan67}, and see an interference pattern emerge in a time 
 $\ll \Gamma^{-1}$ --- before the phase difference between the lasers diffuses to a different 
 random value\footnote{Of course having $\Gamma^{-1}$ longer than a  
 feasible gating time for the detection system is necessary as well, but this is purely a technical requirement. Note that the intensities only have to be stable on this time scale too, so criterion (iv) is not necessary to see this interference.}. 
 The fourth principle is also quantum, and could be argued to be 
 ``more quantum'' in that it requires considering photon-number fluctuations as well as the mean. From his first paper on the topic, 
 quoted above, Glauber regards any light from a thermal source as incoherent, even 
 allowing for ``collimated, completely incoherent beams'' and ``incoherent beams of 
 exceedingly narrow bandwidth''~\cite{Gla63}. For Glauber it is (near) Poissonian statistics 
 that distinguishes a coherent beam from an incoherent one. 
 
 It is worth returning to the opening of the quote from Glauber in Sec.~\ref{sec:intro}, 
 this time in more complete form:\footnote{I have made Glauber's notation more rigorous by replacing $\Pi_k$ with $\bigotimes_k$, for ease of comparison with the rest of this paper.}
 \begin{quote}
The density operator which represents an actual maser beam is not yet known. It is clear that such a beam cannot be represented by a product of individual coherent states, $\bigotimes_k \ket{\alpha_k}\bra{\alpha_k}$,  unless the phase and amplitude stability of the device is perfect. On the other hand, a maser
beam is not at all likely to be described by \dots\ [an]  incoherent classical model \dots. More plausible models for a steady maser beam are much closer in behavior to the 
ideal coherent states.
\end{quote} 
The second sentence, in which $k$ has the same meaning as in the rest of this paper, is undoubtedly true. But while perfect amplitude stability (that is, 
a Poissonian distribution of photon number with a time-constant mean) is a harmless idealisation, perfect phase stability is not. It corresponds to assuming a value of zero 
for the linewidth $\Gamma$. But the fact that a laser (or maser) has 
a minimum value for $\Gamma$ set by quantum fluctuations~\cite{Wis99} 
is arguably the most fundamental result of laser theory,  
discovered by Schawlow and Townes five years before Glauber's paper~\cite{SchTow58}. 
This implies that a CW laser beam is not in a coherent state (or any other pure state), 
so one might wonder whether 
Glauber's terminological contrast of something ``much closer \dots to the ideal coherent states'' 
 with an ``incoherent classical model'' holds up to scrutiny. 

By an ``incoherent classical model''  Glauber means what he called earlier an ``incoherent light beam'',  
defined  (using my notation) as 
$\rho = \bigotimes_k \rho^k_{\rm th}(\nu_k)$. 
This is exactly the same as \erf{cpfts}, but allowing for an arbitrary distribution of mean occupation numbers 
$\nu_k$ rather than my $\nu_k = \nu f(ck)$. This class of states is, in fact (see \ref{sec:apppar}), the {\em only} 
class that are mixtures of coherent states, that describe CW beams, and that can be written as a tensor product 
over different $k$-modes. 
A {\em single realization} of a CW laser beam can certainly be formally written as 
a tensor product of coherent states over different $k$-modes, as in Glauber's expression 
$\bigotimes_k \ket{\alpha_k}\bra{\alpha_k}$, simply by writing the Fourier transform of the 
stochastically varying in space \red{coherent-state amplitude $\alpha(x)$} of \erf{laserscs}. 
However (see \ref{sec:qslb}), there 
seems no way to evaluate the ensemble average to obtain 
the quantum state $\rho$ for a laser --- one has to make do with the infinite ensemble 
of coherent-state realizations itself. 
Thus, even though the idea that laser light is in a coherent state, wrongly attributed to Glauber, 
is indeed wrong, Glauber's distinction --- between incoherent light, for which we can easily write 
an explicit expression for $\rho$ without using the notion of coherent states, 
and a ``steady maser beam'', for which there is no good option but to use coherent 
states in our description --- is still a relevant one.

\section*{Acknowledgments} 
I thank Joe Hope for comments on an earlier draft of the manuscript, which prompted me to revisit 
my (as it turned out, erroneous) faith in \erf{QSLB} and my `derivation' thereof, and 
 Mankei Tsang, Shuming Cheng, Mirko Lobino, 
Andrew White, Dave Kielpinski, Ben Buchler, Geoff Cambpell, Aephraim Steinberg, 
Michael Hall, and Reinhard Werner, 
for comments, mostly on Facebook, that helped me on the just-in-time journey to  
my present (hopefully mostly correct) understanding of the quantum state of a 
laser beam and related topics.  Finally, I sincerely thank Han Woerdman for bringing to my attention earlier work as discussed in the {\bf Note Added} just before the References.

\appendix
\section{Quantum States of CW Beams}

\subsection{The quantum state of a laser beam}\label{sec:qslb}

The simple quantum theory of a laser, pumped far above threshold, predicts a state of 
the cavity mode with the same 
(Poissonian) number statistics as a coherent state~\cite{Wis97,Lou73,SarScuLam74,Lou83,Wis99}. 
It can be modelled as a coherent 
state with fixed amplitude $|\alpha|$ and a phase $\phi$ 
undergoing diffusion, as per \erf{phasebrown}. 
Using stochastic calculus~\cite{IkeWat89}, we can write the phase as a \red{temporally} stochastic variable, 
\beq \label{phasewiener}
\phi(t) = \sqrt{\Gamma}\, W(t),
\eeq
where $W(t)$ is a Wiener process, obeying $\dot{W}(t)=\xi(t)$ with initial condition 
$W(0) = \phi(0)/\sqrt{\Gamma}$.  
That is, the state of the cavity mode can be modelled as a \red{single-mode coherent state 
that varies stochastically in time,}
\beq
\ket{\sqrt{\mu} e^{i\sqrt{\Gamma}\, W(t)}}.
\eeq

Passive optics transforms coherent states into coherent states. Thus with the standard cavity output coupling, 
corresponding to an intensity decay rate for a `cold'  (unpumped) cavity of $\kappa$, 
the output beam can also be described by a tensor product of stochastic coherent states: 
\beq
\ket{\psi} = \bigotimes_j \ket{\sqrt{\mu} e^{i\sqrt{\Gamma}\, W(t_j)}\sqrt{\kappa\Delta t}}^T_{\,j}.
\eeq
Here the $T$ superscript indicates that I am using ``temporal modes'' of duration $\Delta t=t_{j+1}-t_j$, so that $\mu \kappa \Delta t$ is the mean number of photons in each such mode. 
This duration is chosen so that $\omega_0^{-1} \ll \Delta t \ll \Gamma^{-1}$, so that it is infinitesimal on the scale of the stochastic evolution but long enough that we can still use the standard (rotating-wave) treatment for the coupling of the cavity mode to the output modes. 
The time variable here refers to the time at which each bit of the output field is generated 
by this coupling, after which it propagates away at the speed of light. Thus we can equally well 
(in fact, with greater clarity) describe the entire output field, of length $L$ (that is, at a time $L/c$ after the laser reaches steady-state), as a beam with stochastic spatial variation in its phase: 
\beq
\ket{\psi} = \bigotimes_j \ket{\sqrt{\mu} e^{i\sqrt{\Gamma}\, W(x_j/c)}\sqrt{\kappa\Delta x/c}}_{\,j}^X.
\eeq
Here, for mathematical simplicity in converting from time to space, $x=0$ is the point farthest from the laser, and $x=L$ the point at the laser's output mirror. 

The next step is to convert to $k$-space, the reciprocal variable to $x$. 
Since we have a beam of length $L$, $k$ is a discrete variable, 
with $k_{l+1}-k_l = \Delta k = 2\pi/L$ as in Sec.~\ref{sec:coll}. A tensor product of coherent states remains so under any change of mode-basis, and here we have: 
\beq
\ket{\psi} = \bigotimes_l 
 \ket{\sum_j \Delta x\,\sqrt{\mu} e^{i\sqrt{\Gamma}\, W(x_j/c)+ ijl\Delta x\Delta k}\sqrt{\kappa\Delta k / c 2\pi}}_{\,l}^K.
\eeq
Taking the limit $\Delta x \to 0$, 
this becomes, with $k=(2\pi/ L)\, l$ implicit, 
\beq
\ket{\psi}= 
\bigotimes_l 
\ket{\frac{1}{\sqrt{L}}\int_0^L \d x \, e^{ikx}u(x/c)}_{\,l}^K,
\eeq 
where I have defined 
\beq \label{defu}
u(t) = \sqrt{\mu\kappa/c}\,e^{i\sqrt{\Gamma}\, W(t) -i\omega_0t}.
\eeq
Note the equivalence to \erf{laserscs}; 
prior to this point the calculation has been in a frame rotating at the laser frequency $\omega_0$, 
but in \erf{defu} I have changed to a non-rotating frame by explicitly introducing the oscillation frequency. 

Now the overall phase of $u(t)$ is uniformly random, so any explicit expression for the quantum state 
of the laser beam would have no coherences between states with different numbers of photons. 
Thus to try to obtain an explicit expression, like \erf{cpfts}, we should change to the number basis, and consider the projector 
\bqa
\ket{\psi}\bra{\psi} &=& \bigotimes_l \sum_{n,n'} \ket{n}_{\,l}^K\!\bra{n'} \, 
\exp\sq{- \st{\tilde{u}_L(k)}^2} \frac{\ro{\tilde{u}_L(k)}^n 
\ro{\tilde{u}^*_L(k)}^{n'}}{\sqrt{n!}\sqrt{n'!}}\,  ,\label{twoline}
\eqa
where 
\beq \label{tuLK}
\tilde{u}_L(k) \equiv \frac{1}{\sqrt{L}}\int_0^L \d x\, e^{ikx}u(x/c).
\eeq
Now {\em the} quantum state $\rho$ is an ensemble average over all possible realisations: 
$\rho = {\rm E}_W[\ket{\psi}\bra{\psi}]$. To determine this it would be necessary 
to evaluate ensemble averages of products of $\tilde{u}_L(k)$ and 
$\tilde{u}^*_L(k)$ raised to arbitrary non-negative-integer powers, involving simultaneously 
all the possible values of $k=2\pi l/L$. 
There is one simplification: as noted above, the overall phase of $u(t)$ is uniformly random, 
since $\rt{\Gamma}W(0)=\phi(0)$ is uniformly random, and independent of $W(t)-W(0)$. 
Thus the ensemble average will be zero unless we can pair every $\tilde{u}_L(k)$ with a $\tilde{u}^*_L(k')$. 
As the simplest example, 
\bqa \label{sortofspectrum}
{\rm E}[|\tilde{u}_L(k)|^2] &=& \frac{\mu \kappa}{c} \frac{\Gamma/c}{(\Gamma/2c)^2  + (k-k_0)^2}\\
&=& {\nu}\, f(ck),
\eqa
where $f$ is the Lorentzian filter function in \erf{deffom} and ${\nu} = \kappa\mu  (4/\Gamma)$ 
is the number of photons per coherence time, 
as in the body of this paper. The agrees with \erf{firstspectrum}, as expected 
from the Wiener-Khinchin theorem~\cite{IkeWat89}. 

 However, the calculation of the expectation value of more general products seems overwhelming. 
Preliminary explorations suggests that 
the number of (non-stochastic) integrals to be evaluated in the expectation value of a given term 
grows   exponentially with the sum of the powers of the $\tilde{u}$s and $\tilde{u}^*$s in that term. 
There are terms with all possible powers, and, more to the point, a laser beam is a bright beam, with 
of order $\nu = 10^{12}$ photons on average in each $k$-mode. Thus no small-amplitude approximation, 
limiting the size of products that need be considered, is possible.

Perhaps there is a clever trick that would enable $\rho = {\rm E}_W[\ket{\psi}\bra{\psi}]$ to be evaluated, 
but at the present the situation is the same as it was in 1963~\cite{Gla63} with regard to the lack of 
an explicit expression for the quantum state of a laser beam.

\subsection{The quantum state of a thermal-derived beam}\label{qstdb}

In this subsection I show that, despite starting with the same type of equation 
(a stochastically spatially varying coherent state) it {\em is} possible to derive 
an exact expression for the quantum state of a beam if the stochastic coherent-state amplitude 
is as given in \erf{cpftscs}. That is, 
\beq \label{defuth}
u(t) = e^{-i\omega_0t}\sqrt{\nu/c}\int_{-\infty}^{t} \d s\, ({\Gamma}/{2})e^{(\Gamma/2) (s-t)} \zeta(s), 
\eeq
which is a Gaussian random variable (at each $t$). Its Fourier transform,  
\beq \label{tuLKth}
\tilde{u}_L(k) \equiv \frac{1}{\sqrt{L}}\int_0^L \d x\, e^{ikx}u(x/c),
\eeq
is also a set (indexed by $k=2\pi l /L$) of Gaussian random variables, with mean zero, each 
with variance given by \erf{sortofspectrum}. In addition, in the limit of large $L$, 
these variables are independent:\footnote{To obtain this it is necessary to define $u(x)$ to have period $L$, since $k=2\pi l /L, l\in \mathbb{Z}$. This can be done for a stochastic process by selecting, 
from the ensemble of all possible realizations, the subensemble (of measure zero, 
but still continuously infinite) with $u(L/c)=u(0)$. As long as $L \gg c/\Gamma$, 
the local statistical correlations of $u(t)$ everywhere in the interval $[0,L/c]$ are almost unchanged.}
\beq \label{nonast}
{\rm E}[\tilde{u}_L(k)\tilde{u}^*_L(k')] = \delta_{k,k'}\, {\nu}\, f(ck) 
\eeq
Thus each $k$-mode is an independent Gaussian mixture of coherent states with uniformly random phase, 
and mean photon number ${\nu}\, f(ck)$. That is, the quantum state of the beam is 
that of a collimated, polarized, filtered thermal beam (\ref{cpfts}).

\subsection{The non-convergence of the periodogram to the spectrum, and filtering}

The result (\ref{cpfts}) could be derived in the preceding section because of the 
exactly Gaussian statistics of the coherent field, in either $x$- or $k$-space. However, Gaussian 
statistics in the $k$-space are actually generic in the following sense~\cite{PelWu10}: 
if $u(t)$ is a stationary ergodic process, then the limit as $L\to\infty$ of $\tilde{u}_L(k)$ is a 
complex Gaussian random variable of uniformly random phase, whose statistics are thus completely 
defined by its second moment $\nu_k \equiv \lim_{L\to\infty}{\rm E}[|\tilde{u}_L(k)|^2]$. 

This result means that, contrary to what may be found in at least one textbook on stochastic 
methods widely used by physicists~\cite{Gar85}, the periodogram does not converge to the spectrum, 
even for ergodic systems. 
The periodogram is a common way to approximate the spectrum of a stationary stochastic process, 
by taking a long time series of the process, Fourier transforming it, and 
squaring its modulus. But even in the limit of an infinite time series this does not yield the spectrum: 
\beq 
\lim_{L\to\infty}|\tilde{u}_L(k)|^2 \neq S(ck) \equiv \lim_{L\to\infty}{\rm E}[|\tilde{u}_L(k)|^2] 
\eeq
Rather, since $\tilde{u}_L(k)$ is, asymptotically, Gaussian, the periodogram actually has an exponential distribution, 
with a mean, and standard deviation, equal to the spectrum. 

A consequence of the above is that if one were to filter out a single frequency component from any 
infinitely long sample of a stationary stochastic coherent optical field, the 
result would have the same statistics as a single-mode 
thermal state. From shot to shot the measured intensity would differ, according to an exponential distribution. 
Of course in reality there is no way to measure a single frequency component from an infinitely long 
field. However, my presumption is that the above results would hold approximately for a sufficiently narrow filter, 
with width $\delta\omega$ 
much smaller than \red{the spectral width of} the source. 
In this case, the intensity fluctuations ``from shot to shot'' in the filtered field 
would occur on a time scale of order $(\delta \omega)^{-1}$.

In particular, by passing an ideal laser beam through an optical frequency filter with $(\delta \omega) \ll \Gamma$, 
the intensity fluctuations would be greatly increased, from near-Poissonian, with second-order intensity 
autocorrelation function $g^{(2)}(\tau)\approx 1$ for all $\tau$, to massively super-Poissonian, with $g^{(2)}(\tau)$ approaching the thermal value of 2 for $\tau \ll (\delta \omega)^{-1}$. A real laser may have more complicated stochastic phase dynamics than simple diffusion at rate $\Gamma$. It may suffer from  ``technical noise'' such as frequency jitter, equivalent to a narrowband ($\Gamma$) laser whose central frequency wanders around, on a much slower time scale, in a band $\Delta \omega \gg \Gamma$.  
In this case, as the bandwidth $\delta \omega$ of the filter cavity is smoothly reduced I would expect the intensity fluctuations to vary as:
\begin{itemize}
\item near shot-noise, $g^{(2)}\approx 1$, for $\delta \omega \gg \Delta \omega$.
\item well above shot noise, $g^{(2)} > 1$ for $\delta \omega \sim \Delta \omega$.
\item enormous fluctuations (off for most of the time, with bursts of transmission), $g^{(2)}\gg 1$, 
for $\Gamma \ll \delta \omega \ll \Delta \omega$.
\item thermal-like fluctuations, $g^{(2)}\approx 2$, for $\delta \omega \ll \Gamma$.
\end{itemize}

\subsection{Not the quantum state of a laser beam}

As stated in the main text, \erf{QSLB}, here reproduced:
\beq \label{QSLBA}
\bigotimes_k \rho^k_{\rm laser}\left({\nu}f(ck)\right), 
\eeq
is not an approximation  to, or idealisation of, the state of a CW laser beam. 
In fact it does not describe any any CW beam (that is, a beam of indefinite length with 
stationary ergodic statistics). From the preceding subsection, this is simple to see. 
Since a SML state (\ref{mixeq1}) is a mixture of coherent states, so is \erf{QSLBA}. 
 In the spatial mode representation, it is still a mixture of states where each mode is in a
  coherent state. That is, it can be represented by a stochastically varying (in space) 
  coherent-state amplitude. But by the above theorem from Ref.~\cite{PelWu10}, such a 
  stochastic amplitude, if indefinitely extendible and stationary and ergodic (as a CW 
  beam would be) must have an exponential intensity distribution at each frequency, 
  as in a thermal beam. 
  
 If not a laser beam, what then, does \erf{QSLBA} describe? Consider 
 an individual sample, a stochastically varying coherent-state amplitude. This will be a superposition, 
 with random phases, of different spatial frequencies $k$ with deterministic amplitudes 
 $\sqrt{\nu f(ck)}$, creating an irregular pulse of length $L$. In particular, the largest 
 contribution to the pulse is always the $k=k_0$ mode. If we now increase the value of $L$ to $L'$, 
 the nature of a randomly drawn pulse will change. Its largest contribution will now still be the $k=k_0$ mode, 
 but this corresponds to a constant (in the rotating frame) pulse of length $L'$. Thus the 
 state \ref{QSLBA} does not describe a coherent field which is a stationary stochastic {\em process}; 
 it is rather a probabilistic mixture of pulses, each with a shape that depends on the length $L$. 
 That, at least, is my best current understanding. 

\subsection{An apparent paradox} \label{sec:apppar}

For CW beams with a coherent-state description, the results of Ref.~\cite{PelWu10} prove not only that 
$|\tilde{u}_L(k)|^2$ is, asymptotically, exponentially distributed with some mean photon number 
$\nu_k$ (so that each $k$-mode individually is in a SMT state), but that\footnote{Actually, Ref.~\cite{PelWu10} allow $k$ to take all values, not just multiples of $2\pi / L$, and state this result only for almost all pairs $k$, $k'$. 
Thus I am not sure their claim applies to the discrete values I restrict to (as necessary to define orthogonal modes,  
and thus the tensor product Hilbert space $\otimes_k \mathfrak{H}_k$). This may be relevant to the 
existence of the correction term in \erf{corr1L} for non-periodic fields.} 
\beq \label{deltasym}
\lim_{L\to\infty}{\rm E}[\tilde{u}_L(k)\tilde{u}^*_L(k')] = \delta_{k,k'}\, \nu_k,
\eeq
This seems to yield  
the paradoxical conclusion that all such CW beams are what Glauber called incoherent, with each $k$-mode 
independently `prepared' in some SMT state. In particular, since a laser beam can have the same spectrum 
as a thermal-derived beam, it would seem that it would have to be identical in all ways to a thermal-derived 
beam, if it is truly CW. 

The following may point the way to a resolution. For thermal-derived beams, with enforced periodicity, 
the $\delta_{k,k'}$ on the 
right-hand-side of \erf{deltasym} holds for any $L$, as in \erf{nonast}. But for more general stochastic 
coherent fields it appears only in the asymptotic limit. 
For finite but large $L$, there are leading-order corrections scaling as $1/L$. The relevance of such 
terms can be illustrated as follows. If one does not enforce periodicity then, from either the thermal 
model (\ref{defuth}) or the phase-diffusing laser model (\ref{defu}), a relatively simple calculation yields 
the leading order correction 
\bqa \label{corr1L}
{\rm E}[\tilde{u}_L(k)\tilde{u}^*_L(k')] &=& \delta_{k,k'}\, \nu f(ck) \\
&& - 
\nu \frac{2 c}{L\Gamma}f(ck)f(ck')[1-(k'-k_0)(k-k_0)(2c/\Gamma)^2] \nn
\eqa
This might seem like an unimportant correction since we are always interested in the large $L$ limit. 
But it must be remembered that the number of $k$-modes that are significantly populated is of order 
$\Gamma/(c\Delta k) = L \Gamma / (2\pi c)$. Thus the small correction for each pair of modes 
sums, over all modes, to something non-negligible. This is how it is possible to obtain, 
when taking the Fourier transform back to $x$-space, a field $u(x/c)$ with {\em only} local correlations, 
and no artificial high correlations between $u$ for $x < c/\Gamma$ and $u$ for $x > L - c/\Gamma$. 

The above correction has nothing to do with non-Gaussianity; as stated above, it appears in the thermal 
case for a non-period field. But it does illustrate how the theorem stating that the periodograms $\tilde{u}_L(k)$ 
for different $k$ are asymptotically pairwise independent Gaussians could be compatible with the 
field $u$ itself being strongly non-Gaussian, as in the phase diffusion model (\ref{defu}). If the independence 
between different $k$ were exact, with no corrections scaling as $1/L$, then 
the coherent-state amplitudes in the different frequency modes would be 
strictly statistically independent, and the only allowed CW beams would be Glauber's ``incoherent light beams'' with 
$\rho = \bigotimes_k \rho^k_{\rm th}(\nu_k)$. 
 
 \section*{Note Added in Proof}
Subsequent to acceptance it was brought to my attention that the `prediction' I made in Sec.~4.4 (see also Appendix~A.3), that a laser filtered well within its linewidth would be indistinguishable from a thermal beam, is by no means new. In essence it was proven by Armstrong in 1966~\cite{Arm66}, although he found the negative-exponential intensity distribution characteristic of thermal sources only numerically, not analytically. Moreover, an experiment seeing an increase in $g^{(2)}(0)$ towards $2$, as the filter was narrowed, was performed in 1992~\cite{Nee92}. See also Ref.~\cite{Nee93}.

\section*{References}
\bibliography{How_many_principles-6-post-proof.bbl}

\begin{thebibliography}{10}

\bibitem{wikiLaser}
Laser.
\newblock \url{https://en.wikipedia.org/wiki/Laser}, 1st August 2015.

\bibitem{PicWil65}
R.~H. Picard and C.~R. Willis.
\newblock Coherence in a model of interacting radiation and matter.
\newblock {\em Phys. Rev.}, 139:A10--A15, Jul 1965.

\bibitem{Gla63}
Roy~J. Glauber.
\newblock Photon correlations.
\newblock {\em Phys. Rev. Lett.}, 10:84--86, Feb 1963.

\bibitem{iyolLaser}
Lasers.
\newblock \url{http://www.light2015.org/Home/LearnAboutLight/Lasers.html}, 1st
  August 2015.

\bibitem{Wis97}
H.~M. Wiseman.
\newblock Defining the (atom) laser.
\newblock {\em Phys. Rev. A}, 56:2068--2084, Sep 1997.

\bibitem{Lou73}
W.~H. Louisell.
\newblock {\em Quantum Statistical Properties of Radiation}.
\newblock John Wiley \& Sons, New York, 1973.

\bibitem{SarScuLam74}
M.~Sargent, M.~O. Scully, and W.~E. Lamb.
\newblock {\em Laser Physics}.
\newblock Addison-Wesley, Reading, Mass., 1974.

\bibitem{Lou83}
R.~Loudon.
\newblock {\em The Quantum Theory of Light}.
\newblock Oxford University Press, Oxford, 1983.

\bibitem{Wis04a}
Howard~M. Wiseman.
\newblock Defending continuous variable teleportation: why a laser is a clock,
  not a quantum channel.
\newblock {\em Journal of Optics B: Quantum and Semiclassical Optics},
  6(8):S849, 2004.

\bibitem{WhiteDwarf}
K.~Werner, T.~Rauch, and J.~W. Kruk.
\newblock Discovery of photospheric {CaX} emission lines in the far-{UV}
  spectrum of the hottest known white dwarf ({KPD 0005+5106}).
\newblock {\em Astronomy \& Astrophysics Letters}, 492:L43, 2008.

\bibitem{Hecht01}
E.~Hecht.
\newblock {\em Optics (4th ed.)}.
\newblock Addison-Wesley, Reading, Mass., 2001.

\bibitem{IkeWat89}
N.~Ikeda and S.~Watanabe.
\newblock {\em Stochastic Differential Equations and Diffusion Processes; 2nd
  ed.}
\newblock North-Holland Publishing Co., Amsterdam, 1989.

\bibitem{PflMan67}
R.~L. Pfleegor and L.~Mandel.
\newblock Interference of independent photon beams.
\newblock {\em Phys. Rev.}, 159:1084--1088, Jul 1967.

\bibitem{Wis99}
H.~M. Wiseman.
\newblock Light amplification without stimulated emission: Beyond the standard
  quantum limit to the laser linewidth.
\newblock {\em Phys. Rev. A}, 60:4083--4093, Nov 1999.

\bibitem{SchTow58}
A.~L. Schawlow and C.~H. Townes.
\newblock Infrared and optical masers.
\newblock {\em Phys. Rev.}, 112:1940--1949, Dec 1958.

\bibitem{PelWu10}
Magda Peligrad and Wei~Biao Wu.
\newblock Central limit theorem for {F}ourier transforms of stationary
  processes.
\newblock {\em Ann. Probab.}, 38(5):2009--2022, 09 2010.

\bibitem{Gar85}
C.~W. Gardiner.
\newblock {\em Handbook of Stochastic Methods}.
\newblock Spring\-er, Berlin, 1985.

\bibitem{Arm66}
J.~A. Armstrong.
\newblock Theory of interferometric analysis of laser phase noise$\ast$.
\newblock {\em J. Opt. Soc. Am.}, 56(8):1024--1031, Aug 1966.

\bibitem{Nee92}
R.~Centeno~Neelen, D.~M. Boersma, M.~P. van Exter, G.~Nienhuis, and J.~P.
  Woerdman.
\newblock Spectral filtering within the schawlow-townes linewidth of a
  semiconductor laser.
\newblock {\em Phys. Rev. Lett.}, 69:593--596, Jul 1992.

\bibitem{Nee93}
R.~Centeno Neelen, D.~M. Boersma, M.~P. van Exter, G.~Nienhuis, and J.~P.
  Woerdman.
\newblock Spectral filtering within the schawlow-townes linewidth as a
  diagnostic tool for studying laser phase noise.
\newblock {\em Optics Communications}, 100:289 -- 302, 1993.

\end{thebibliography}


\end{document}